\documentstyle[12pt]{article}

\begin{document}
\newcommand {\be}{\begin{equation}}
\newcommand {\ee}{\end{equation}}
\newcommand {\bea}{\begin{array}}
\newcommand {\cl}{\centerline}
\newcommand {\eea}{\end{array}}
\baselineskip 0.65 cm
\begin{flushright}
IPM-96-174\\
hep-th/9612056
\end{flushright}
\begin{center}
\bf \Large {D-brane Interactions, World-sheet Parity and Anti-Symmetric Tensor}
\end{center}
\begin{center}
H. Arfaei\footnote{e-mail: Arfaei@ theory.ipm.ac.ir} 
$\;\;\;\;\;$ and $\;\;\;\;$
M.M. Sheikh Jabbari\footnote{e-mail: jabbari@netware2.ipm.ac.ir}
\end{center}
\cl {\it Institute For Studies in Theoretical Physics and Mathematics 
IPM} \cl{\it P.O.Box 19395-5531, Tehran,Iran}
\cl{\it Department of Physics Sharif University of Technology}
\cl{\it P.O.Box 11365-9161, Tehran, Iran}
\vskip 2cm
\begin{abstract}
 Using world-sheet parity we show that mixed D and N components of D-strings
are dual to the anti-symmetric $B_{\mu\nu}$ field. The contribution of the latter 
is responsible for the reduction and even removal of all the interactions between two dissimilar 
D-branes.
\end{abstract}
\newpage
 In the past year the study of D-branes and D-strings has attracted much 
attention since they may be used to obtain insight into non-perturbative 
structure underlying the strings. They have shown the properties one expects from 
the extended solutions to field thoeretic low energy limit of the (type-II) 
string theories mainly at long distances. It was through this kind of study 
that $RR$ charge of D-branes hidden in the simple
topological defect, the brane, were discovered [1,2].

 Attention has been concentrated mainly on the $RR$ charges, which the interaction
of similar D-branes reveals it. A particular $D_p$-brane of dimension $p$ can emit
and absorb quanta of the $RR\ p+1 $ form. Therefore consideration of similar D-brane
interactions manifests their $RR$-charge. On the other hand consideration of 
dissimilar branes hides their $RR$-charge and may reveal other properties hidden
by similarity [3]. 

 Recently we have found [4] that in certain conditions the gravitational attraction
of D-branes can be balanced or even reversed. In this note we will show that
the balancing force is due to the exchange of anti-symmetric $B_{\mu\nu}$ tensor
which doesn't participate  in the interaction of similar branes because they form
symmetric states. 

 To extract this information we exploit the anti-symmetric nature of the amplitude under
world-sheet parity. This is done by insertion of the appropriate operator in the
corresponding trace for the calculation of the amplitude. 

 We will show that the $DN$ modes of the string stretched between branes are 
dual to the $B_{\mu\nu}$ tensor fields of closed channel. The dependence of the 
effect of $B_{\mu\nu}$ on the relative angle and velocity is also discussed which
will give us a handle for probing short distance behaviour of D-branes.

 We impose the proper boundary conditions on the string with the ends on two 
parallel branes of dimensions $p,p'$ as given in [4]. The boundary conditions
on the fermionic part is induced by world-sheet SUSY. We have two kinds of open D-strings $NS$ and $R$ type. 
The mode expansions of the bosonic and fermionic varibles are:

\be
\begin{array}{cc}
NN\ components\ \  X^{\mu}=p^{\mu}\tau +\sum_{n \in Z} {1\over n} \alpha_n^{\mu} e^{-in\tau} \cos n\sigma &  \mu= 0,...,p' \\
DN\ components \ \ \ \ \ \ =\sum_{r \in Z+1/2} {1\over r} \alpha_r^{\mu} e^{-ir\tau} \sin r\sigma &  \mu=p'+1,...,p  \\      
DD\ components \ \ \ \ \ =Y^{\mu} {\sigma\over \pi}+\sum_{n \in Z} {1 \over n} \alpha_{n}^{\mu}e^{-in\tau} \sin n\sigma &  \mu=p+1,...,9. 
\end{array}
\ee
for both $R$ and $NS$ types, and for the $\psi_{\pm}^{\mu}$, world-sheet fermions:                           
\be
\left\{ \bea{cc}
\psi^{\mu}_+=\sum_{n \in Z} d^{\mu}_n e^{-in(\tau+\sigma)} \;\;\; 
\psi^{\mu}_-=\sum_{n \in Z} d^{\mu}_n e^{-in(\tau-\sigma)} \;\;\;\mu=0,...,p'  \\
\psi^{\mu}_+=\sum_{n \in Z} d^{\mu}_n e^{-in(\tau+\sigma)} \;\;\; 
\psi^{\mu}_-=-\sum_{n \in Z} d^{\mu}_n e^{-in(\tau-\sigma)} \;\;\;\mu=p+1,...,9 \\
\psi^{\mu}_+=\sum_{r \in Z+1/2} d^{\mu}_r e^{-ir(\tau+\sigma)} \;\;\; 
\psi^{\mu}_-=-\sum_{r \in Z+1/2} d^{\mu}_r e^{-ir(\tau-\sigma)} \;\;\;\mu=p'+1,...,p 
\eea \right.
\ee
for the {\bf{R}}-sector and
\be
\left\{ \bea{cc}
\psi^{\mu}_+=\sum_{r \in Z+1/2} b^{\mu}_r e^{-ir(\tau+\sigma)} \;\;\; 
\psi^{\mu}_-=\sum_{r \in Z+1/2} b^{\mu}_n e^{-ir(\tau-\sigma)} \;\;\;\mu=0,...,p'  \\
\psi^{\mu}_+=\sum_{r \in Z+1/2} b^{\mu}_r e^{-ir(\tau+\sigma)} \;\;\; 
\psi^{\mu}_-=-\sum_{r \in Z+1/2} b^{\mu}_n e^{-ir(\tau-\sigma)} \;\;\;\mu=p+1,...,9 \\
\psi^{\mu}_+=\sum_{n \in Z} b^{\mu}_n e^{-in(\tau+\sigma)} \;\;\; 
\psi^{\mu}_-=-\sum_{n \in Z} b^{\mu}_n e^{-in(\tau-\sigma)} \;\;\;\mu=p'+1,...,p 
\eea \right.
\ee
for the {\bf{NS}}-sector.

 Following the general approach for studying the interaction of D-branes [2] we consider
a world-sheet corresponding to the exchange of a closed strings between the branes
which is the same as a loop of D-string in the crossed channel. The focus of our
attention is on the world-sheet parity . From the closed string point of view the world-sheet parity
which can be introduced by $\ \sigma \leftrightarrow \pi-\sigma\ $ or equivalently
$\ z \leftrightarrow -\bar z\ $, interchanges left moving and right moving modes.

The graviton and dilaton and $RR$ fields are even under this transformations, 
but $B_{\mu\nu}$ (anti-symmetric tensor field) is odd [2]. 
 On the other hand $\ z \leftrightarrow -\bar z\ $ also acts as world-sheet parity
on the open D-strings such that

\be
\left\{ \bea{cc}
 Y^{\mu} \rightarrow -Y^{\mu}  \;\;\;\; ;\;\;\;  p^{\mu} \rightarrow  p^{\mu}  \\
 \alpha^{\mu}_n \rightarrow  (-)^n \alpha^{\mu}_n \;\;\; ;\;\;\; \alpha^{\mu}_r \rightarrow  (-)^{r-1/2} \alpha^{\mu}_r.
 \eea \right.
\ee
 As is easily seen $DD$,$NN$ components of D-strings are even and $DN$ components are odd.

 In the case of two similar parallel D-branes the system is symmetric under world-sheet parity and therefore
the $B_{\mu\nu}$ field doesn't contribute  to the interactions, but when the D-branes
are of different dimensions or not-parallel the  $DN$ components of the open strings are present which
give rise to a non-world-sheet parity invariant contribution manifested as the 
exchange of anti-symmetric $B_{\mu\nu}$ field in the closed channel. These observations
are sufficient to show that the $DN$ modes in open string channel and anti-symmetric
tensor field in the closed string channel are dual to each other.

\ \ \ \ They represent the contribution to amplitude originating from the odd fields under 
$\ z \leftrightarrow -\bar z\ $.

 To make the point more clear we can guage the thoery with respect to world-sheet parity
transformation which removes all the odd fields and keeps only the invariant(even) parts. 

 In the following we will use the above consideration to represent the world-sheet parity
odd and even contributions to the amplitude and show that the  share of graviton-dilaton and 
$B_{\mu\nu}$ enter with opposite sign.

 The amplitude for the interaction in our discussions is given by: 
\be 
A=\int {dt \over 2t} \sum_{i,p}e^{-2\pi\alpha' t (p^2+M_i^2)}
\ee 
where $i$ indicates the modes of the open string and $p$ indicate momentum which has non-zero
components in the first $p'+1$ dimensions.

 In order to calculate the "Trace" we have to insert the corresponding GSO projcetions for
$NS$, $R$ sectors separately to remove the tachyon contribution from the closed string
channel. We can also introduce the world-sheet parity projection which separates
$B_{\mu\nu}$ and $\{G_{\mu\nu},\Phi\}$ (odd and even) contributions.
 Doing so we obtain the amplitude $A=A_+ +A_- $, 
\be
 A=V_{p'+1} \int \frac{dt}{2t}(8\pi^2\alpha't)^{-(p'+1)/2}e^{-\frac{Y^2t}{2\pi^2\alpha'}} 
 (\bf{ NS_{\pm}-R_{\pm}}). 
\ee
where $\bf{NS_{\pm}}$ and $\bf{ R_{\pm}}$ are given by 
\be
{\bf NS_{\pm}} = 2^{-2}\ ({f_1 \over f_3})^{-8+\Delta}\bigl[({f_2 \over f_4})^{\Delta}
 \pm ({f_3 \over f_4})^{\Delta} \bigr]
\ee
\be
{\bf R_{\pm}}= 2^{-2}\ ({f_1 \over f_2})^{-8+\Delta}\bigl[({f_3 \over f_4})^{\Delta}
 \pm ({f_2 \over f_4})^{\Delta} \bigr]
\ee
where $q=e^{-\pi t}$ and $\pm$ stands for positive and negative parity of exchanged
closed string respectively and $f_i$ are given in terms of $\Theta$ functions by:

\be
\bea{cc} 
 f_1=({\Theta_1' \over 2\pi})^{1/3} \;\;\;\; ;\;\;\;  f_2=f_1 ({2\pi \Theta_2 \over \Theta_1'})^{1/2} \\

 f_3=f_1 ({2\pi \Theta_3 \over \Theta_1'})^{1/2} \;\;\;\; ;\;\;\;\;  f_4=f_1 ({2\pi \Theta_4 \over \Theta_1'})^{1/2} 

\eea 
\ee
 Before extracting the contribution of the dilaton and graviton and $B_{\mu\nu}$
we note that when the number of symmetric D-string components and anti-symmetric 
components are equal the amplitude vanishes exactly. This occurs when
\be
 p'+(8-p)=p-p' \;\;\; ;\;\;\ or  \;\;\;\; \Delta=4.
\ee
  As is shown in [4] the cancellation takes place order by order in all powers of $t$.

  Taking the small $t$ limit of the integrand we can separate the massless 
closed string, the dilaton and graviton and $B_{\mu\nu}$, 
contributions which show the long range interactions: 
\be
A_{B_{\mu\nu}} ={-1 \over 4} 2^{\Delta}V_{p'+1}(4\pi^2\alpha')^{3-{p+p' \over 2}}\ {\Delta}\pi G_{9-p}(Y^2),
\ee
\be
A_{G,\Phi} ={1 \over 4}2^{\Delta} V_{p'+1}(4\pi^2\alpha')^{3-{p+p' \over 2}}\ {(8-\Delta)}\pi G_{9-p}(Y^2).
\ee
 It is worth noting that since the branes are of different dimensions there is no
$RR$ interaction, besides the $B_{\mu\nu}$ charge density of the D-branes is the same as their mass density
and equal to $(4\pi^2\alpha')^{(3-p)/2}$. 

 Similar analysis works for branes at non-zero angle considered in [4]. For these branes
complete cancellation occurs at  $\Delta=2,\theta=1/2$
(p,p+2 perpendicular D-branes). Again in this case we have four world-sheet parity
symmetric components and four anti-symmetric components. Although $\Delta=2$
there are two other anti-symmetric components having their origin in the angle  
$\pi/2$ between  the branes.

 Since the cancellation takes place in all orders of $t$, we can smoothly take
$Y$, the distance between branes, to be zero and obtain a composite brane with two different
$RR$ fields, dilaton and graviton and $B_{\mu\nu}$ field [5,6,7,8,9,10,11], one consisting of
two perpendicular intersecting parts with $\Delta=2$ and the other consisting of a
p-brane with a $(p-4)$-brane sitting in it.

 Similar results could be obtained in low energy limit from $DBI$ action which  
governs D-brane low energy dynamics (it only concerns the exchange of massless fields  
between D-branes.)

 $DBI$ action for bosonic varibles is given by [12,13]:
\be
  S=T_p \int d^{p+1} \eta \sqrt{det (h+\cal F)}
\ee
 where $\eta$ is the world-brane coordinates and
\be
h_{ab}=g_{\mu\nu} \frac {\partial{X^{\mu}}}{\partial{\eta^a}}\frac {\partial{X^{\nu}}}{\partial{\eta^b}}   
\ee
is the induced metric on the brane. $\cal F$ is the $U(1)$ gauge field strength: 
\be
{\cal F}= B_{\mu\nu}\frac {\partial{X^{\mu}}}{\partial{\eta^a}}\frac {\partial{X^{\nu}}}{\partial{\eta^b}}-F_{ab}  \;\;\;\;\; ;\;\;\ F=dA.
\ee
 The action in  the flat back ground for weak field limit could be expanded as:
\be
S=-T_p \int_{brane} Tr {\cal F}^2 + O({\cal F}^3)
\ee
which is the usual $U(1)$ gauge action. In order to study the dynamics of D-brane
and $B_{\mu\nu}$ we must add the kinetic term of $B_{\mu\nu}$ to this action [11]:
\be
S=\int_{\cal M} \mid dB \mid ^2-T_p \int_{brane} Tr {\cal F}^2 
\ee 
where $\cal M$ is the space-time manifold, and the brane as a flat submanifold of it.

 By using reparametrization invariance for $\eta$,the coordinates $X^\mu$ (space-time 
 coordinates) could be chosen as:
\be
X^\mu =\eta^a \delta^{\mu}_a +\Phi^A \delta^{\mu}_A
\ee
where $a=1,...,p$ and $A=p+1,...,D=dim {\cal M}$. $\Phi^A\mid_{brane}=\Phi^A(\eta)$,    
are scalars respect to the brane, showing D-brane transverse fluctuations. $A_a(\eta)$
is $U(1)$ gauge field which represents the internal modes of brane. 

 In terms of this new coordinates the action (17) up to $O(\Phi^2)$ is written as:
\be
S=\int_{\cal M} \mid dB \mid ^2-T_p \int_{brane} (B_{ab}-F_{ab})^2 -2T_p \int_{brane} (B_{ab}-F_{ab})B_{aA}{\partial_b \Phi^A}.
\ee

 Hence $A_a$ and $B_{\mu\nu}$ field equations are (by using gauge freedom):
\be
\left\{ \bea{cc}

 \Box_D B_{\mu\nu}=T_p {\cal F}_{ab}\ \delta^a_{\mu}\ \delta^b_{\nu}\ \delta(bran0

 \partial_a {\cal F}_{ab}=0
\eea \right.
\ee 
  As we see ${\cal F}_{ab}$ acts as a $B_{\mu\nu}$ source(charge) and also the 
action written above gives the $B_{\mu\nu}$ interaction of D-branes, 
which comes from the third term of action(the second term gives no contribution to
interaction). This term vanishes in the case of two similar branes.It is proportional to number of 
DN modes because of $B_{aA}$ term. The same argument is held when we cosider gravity 
(non-flat metric). So D-brane has $B_{\mu\nu}$ charge because of both its internal modes
($F_{ab}$) and also its transverse fluctuations($\partial_a \Phi^A$).

 In string theoretic point of view these scalars and $U(1)$ fields can be intrepreted as D and N 
components of D-strings respectively which are attached to one D-brane. As it is
discussed in [11] D-strings (or D-branes) are $B_{\mu\nu}$ charged.

 Extension of our analysis to moving branes shows that when branes have relative
velocity $v$  exchange of $B_{\mu\nu}$ and graviton both give the non-zero velocity dependent potential [4]:

\be
 V_{NSNS}=-2V_0 (\frac {4-\Delta-v^2 (2-\Delta)}{1-v^2} ) \;\;\;\; ;\;\; V_0=2^{\Delta}V_{p'} (4\pi^2 \alpha')^{3-(p+p')/2}G_{9-p}(Y^2).
\ee

which in $\Delta=4$ non-relativistic limit it's proportional to $v^2$, and potential is attractive.
By parity arguments it is easily seen that graviton and dilaton 
contribution is proportional to $\frac{1-v^2/2}{1-v^2}$. This point
can be used to probe $B_{\mu\nu}$ behaviour of $D_p$-branes by $(p-4)$ or
$(p-2)$ D-branes [14,15].

 The particular cases of interest are: 

In type IIA theory

 $D_0$ and $D_4$ branes 

 $D_2$ and $D_4$ branes  perpendicular 

 $D_2$ and $D_6$ branes 

 $D_4$ and $D_6$ branes  perpendicular 

In type IIB: 

 $D_1$ and $D_5$ branes   

 $D_1$ and $D_3$ branes at $\theta=\pi/2$  

 $D_3$ and $D_5$ branes at $\theta=\pi/2$   

  SUGRA solutions with some of the above configurations have been found and we 
 conjecture that the others also exist as p-brane solutions. Work in this direction 
 is being pursued.

{\bf Acknowledgement:}

Authors would like to thank Farhad. Ardalan for fruitful discussions.

\end{document}